\documentclass[aps,prl,showpacs]{revtex4}
\usepackage{graphicx}
\usepackage{amsfonts,amssymb,amsmath}
\usepackage{bm}

\begin{document}

\title{Symmetries of the Jahn Teller System and their Solvability}
\date{\today}
 
\author{Eser K\"{o}rc\"{u}k}
\email{korcuk@gantep.edu.tr}
%\homepage{www.mrl.ucsb.edu/~ederer} 
\affiliation{Department of Physics, Faculty of Engineering 
University of Gaziantep,  27310 Gaziantep, Turkey}
\author{Ramazan Ko\c{c}}
\email{koc@gantep.edu.tr}
\affiliation{Department of Physics, Faculty of Engineering 
University of Gaziantep,  27310 Gaziantep, Turkey}
\begin{abstract}
We present a method of obtaining the quasi exact solution of 
the Jahn Teller systems in the framework of osp(2,2) superalgebra. 
The hamiltonian have been solved in the Bargmann-Fock space by obtaining an 
expression as linear and bilinear combinations of the generators of osp(2,2).  
In particulare, we have discussed quasi exact solvability of $E\times \varepsilon $  Jahn-Teller 
Hamiltonian.
\end{abstract}

\maketitle

\section{Introduction}

The Jahn-Teller (JT) distortion problem is an old one, dating back over
sixty years \cite{1}. Yet, even today, new contributions to this problem are
being made \cite{3}. They appear, however, not to have been fully exploited
in the analysis of JT problem. The $E\otimes \varepsilon $ JT problem is a
system with doubly degenerate electronic state and doubly degenerate JT
active vibrational state. The JT effect describes the interaction of
degenerate electronic states through non-totally symmetric, usually
non-degenerate, nuclear modes. This effect plays an important role in
explaining the structure and dynamics of the solids and molecules in
degenerate electronic states.

The studies of the JT effect led Judd to discover a class of exact isolated
solutions of the model \cite{3}. The complete description of these solutions
have been given by Reik et al\cite{5}. They observed that the isolated
solutions could be obtained by using Neumann series of expansions of the
eigenvectors in the Bargmann-Fock space described by the boson operators.
The same problem has been treated in\cite{6}.

On the other hand, the conception of quasi-exactly solvable (QES) systems
discovered \cite{9,12,13,14} in the 1980's, has received much attention in
recent years, both from the viewpoint of physical applications and their
inner mathematical beauty. The classification of the $2\times 2$ matrix
differential equations in one real variable possessing polynomial solution
have been described \cite{11,15}. The relevant algebraic structure of the $%
E\otimes \varepsilon $ JT system is the graded algebra $osp(2,2)$ and in
this poster, we present a quasi-exact solution of the $E\otimes \varepsilon $
JT Hamiltonian.

\section{Symmetry Properties of the $E\times \protect\varepsilon $ Jahn
Teller System}

In this section a group theoretical treatment of \ JT distortion, in general
case of two-fold degenerate states of various groups is provided. The JT
interaction matrices and surface energies have been obtained by using
symmetry properties of the system\cite{rh}. Let us start by describing the
Hamiltonian that generates $D^{\ell }\otimes D^{\ell }$ surface, where $%
D^{\ell }$ denotes the irreducible representation. The standard Hamiltonian
may be written in the form%
\begin{equation}
H=H_{0}+H_{JT}  \label{r1}
\end{equation}%
where $H_{0}$ describes free (uncoupled) electron/holes and their
vibrational states and $H_{JT}$ is the Jahn-Teller interaction Hamiltonian.
It is known that the Hamiltonian of the JT coupling is invariant under the
rotational operations of the $SO(3)$ group. The totally symmetric part of
direct product of an irreducible representations of a finite group, which
describes the properties of the JT system can be expressed in the form of%
\begin{equation}
\left[ D^{\ell }\otimes D^{\ell }\right] =D^{\ell _{1}}\oplus D^{\ell
_{2}}\oplus \cdots \oplus D^{\ell _{n}}  \label{r2}
\end{equation}%
where $\ell $ is the angular momentum quantum number. Decomposition of $%
\left[ D^{\ell }\otimes D^{\ell }\right] $ implies that the JT Hamiltonian
can be written in the following way%
\begin{equation}
H_{JT}=H^{\ell _{1}}+H^{\ell _{2}}+\cdots H^{\ell _{n}}  \label{r3}
\end{equation}%
where $H^{\ell _{i}}$ is the JT Hamiltonian and its invariant under the
symmetry operations of the corresponding finite group, for the $2\ell +1$
dimensional representation. As an example consider an icosahedral symmetric
system. The symmetric part of the $H_{g}$ interaction is given by%
\begin{equation}
\left[ H_{g}\otimes H_{g}\right] =A_{g}\oplus H_{g}\oplus (H_{g}\oplus G_{g})
\label{r4}
\end{equation}%
where $A_{g},H_{g},$ and $G_{g}$ are the irreducible representations of the
icosahedral group $I_{h}$. Since $I_{h}$ is a subgroup of $O(3)$
decomposition of the coupling of the $\ell =2$ state can be written as%
\begin{equation}
\left[ D^{2}\otimes D^{2}\right] =D^{0}\oplus D^{2}\oplus D^{4}  \label{r5}
\end{equation}%
and its Hamiltonian is given by%
\begin{equation}
H_{JT}=H^{0}+H^{2}+H^{4}.  \label{r6}
\end{equation}%
The Hamiltonians $H^{i}$ must be separately invariant under the symmetry
group $I_{h}$. The symmetric part contains the totally symmetric
representation $H^{0}=A_{g}$ can exactly be solved. Before going further we
list the decomposition of the symmetric products of the $\left[ E\otimes E%
\right] $ JT interaction and corresponding symmetry groups%
\begin{eqnarray*}
O_{h} &:&\left[ E\otimes E\right] =A_{1g}\oplus E;\quad T_{h}:\left[
E\otimes E\right] =A_{g}\oplus E \\
D_{2p} &:&\left[ E\otimes E\right] =A_{1}\oplus E;\quad C_{2p}:\left[
E\otimes E\right] =A_{1}\oplus E
\end{eqnarray*}%
where $O_{h},T_{h},D_{2p}$ and $C_{2p}$ denotes octahedral, tetrahedral,
dihedral and cyclic groups, respectively. In the following section we
discuss the construction of the $E\otimes \varepsilon $ JT Hamiltonian.

\section{The $E\otimes \protect\varepsilon $ Jahn-Teller Hamiltonian}

The well-known form of the $E\otimes \varepsilon $ JT Hamiltonian describing
a two-level fermionic subsystem coupled to two boson modes has been given by
Reik\cite{5}.%
\begin{equation}
H=a_{1}^{+}a_{1}+a_{2}^{+}a_{2}+1+(\frac{1}{2}+2\mu )\sigma _{0}+2\kappa
\lbrack (a_{1}+a_{2}^{+})\sigma _{+}+(a_{1}^{+}+a_{2})\sigma _{-}]
\label{eq:1}
\end{equation}%
where $\frac{1}{2}+2\mu $ is the level separation, $\kappa $ is the coupling
strength. The Pauli matrices $\sigma _{\pm ,0\text{ }}$are given by%
\begin{equation}
\sigma _{+}=\left[ 
\begin{array}{cc}
0 & 1 \\ 
0 & 0%
\end{array}%
\right] ,\quad \sigma _{-}=\left[ 
\begin{array}{cc}
0 & 0 \\ 
1 & 0%
\end{array}%
\right] ,\;\sigma _{0}=\left[ 
\begin{array}{cc}
1 & 0 \\ 
0 & -1%
\end{array}%
\right] .  \label{eq:2}
\end{equation}%
The annihilation and creation operators, $a_{i}\;$and$\;a_{i}^{+}$ satisfy
the usual commutation relations%
\begin{equation}
\lbrack a_{i}^{+},a_{j}^{+}]=[a_{i},a_{j}]=0,\quad \lbrack
a_{i},a_{j}^{+}]=\delta _{ij}.  \label{eq:4}
\end{equation}%
The number operator of the Hamiltonian (\ref{eq:1}), $J_{1},$ represents the
angular momentum of the system and is given by%
\begin{equation}
J_{1}=a_{1}^{+}a_{1}-a_{2}^{+}a_{2}+\frac{1}{2}\sigma _{0}.  \label{eq:6}
\end{equation}%
Note that $J_{1}$ commutes with $H$ and the eigenvalue problem of the
angular momentum part can be easily solved and it reads%
\begin{equation}
J_{1}\left| \psi \right\rangle _{j+\frac{1}{2}}=\left( j+\frac{1}{2}\right)
\left| \psi \right\rangle _{j+\frac{1}{2}}\quad j=0,1,2\cdots  \label{eq:7}
\end{equation}%
with the eigenfunctions%
\begin{equation}
\left| \psi \right\rangle _{j+\frac{1}{2}}=(a_{1}^{+})^{j}\phi
_{1}(a_{1}^{+}a_{2}^{+})\left| 0\right\rangle \left| \uparrow \right\rangle
+(a_{1}^{+})^{j+1}\phi _{2}(a_{1}^{+}a_{2}^{+})\left| 0\right\rangle \left|
\downarrow \right\rangle .  \label{eq:8}
\end{equation}%
where $\left| 0\right\rangle $ is the vacuum state for both bosons. Here $%
\left| \uparrow \right\rangle $ and $\left| \downarrow \right\rangle $ are
the eigenstates of the $\sigma _{0}$, $\phi _{1}$ and $\phi _{2}$ are
arbitrary functions of $a_{1}^{+}a_{2}^{+}$ . Because the operators $H$ and $%
J_{1}$ commute, the eigenfunctions (\ref{eq:8}) are also the eigenfunctions
of the Hamiltonian (\ref{eq:1}).Therefore we can write the eigenvalue
equation,%
\begin{equation}
H\left| \psi \right\rangle _{j+\frac{1}{2}}=E\left| \psi \right\rangle _{j+%
\frac{1}{2}};\quad E=2\epsilon +j+\frac{3}{2}.  \label{eq:9}
\end{equation}%
The Hamiltonian $H$ can be expressed in the Bargmann-Fock space by using the
realizations of the bosonic operators%
\begin{equation}
a_{i}^{+}=z_{i},\quad a_{i}=\frac{d}{dz_{i}},\;i=1,2.  \label{eq:12}
\end{equation}

In this formulation, the Hamiltonian $H$ consists of two independent sets of
first order linear differential equations. Substituting (\ref{eq:8}) and (%
\ref{eq:12}) into (\ref{eq:1}) and defining $\xi =z_{1}.z_{2}$ one can
obtain the following two linear differential equations satisfied by the
functions $\phi _{1}$ and $\phi _{2}$: 
\begin{subequations}
\begin{equation}
\lbrack \xi \frac{d}{d\xi }-(\epsilon -\mu )]\phi _{1}+\kappa \lbrack \xi 
\frac{d}{d\xi }+(\xi +j+1)]\phi _{2}=0;\quad \kappa \lbrack \frac{d}{d\xi }%
+1]\phi _{1}+[\xi \frac{d}{d\xi }-(\epsilon +\mu )]\phi _{2}=0.
\label{eq:13b}
\end{equation}

These coupled differential equations represent the Schr\"{o}dinger equation
of the $E\otimes \varepsilon $ JT system in Bargmann's Hilbert space and its
isolated exact solution have been obtained by Reik\cite{5}. In this poster
we follow a different strategy to solve the Hamiltonian(\ref{eq:1}) and we
show the Hamiltonian possesses $osp(2,2)$ symmetry.

\section{Two-boson one fermion $osp(2,2)$ superalgebra}

In order to construct $osp(2,2)$ superalgebra let us start by introducing
three generators of the $su(1,1)$ algebra, 
\end{subequations}
\begin{equation}
J_{+}=a_{1}^{+}a_{2}^{+},\quad J_{-}=a_{2}a_{1},\quad J_{0}=\frac{1}{2}%
\left( a_{1}^{+}a_{1}+a_{2}^{+}a_{2}+1\right)  \label{1}
\end{equation}%
These are the Schwinger representation of $su(1,1)$ algebra and its number
operator is given by,%
\begin{equation}
N=a_{1}^{+}a_{1}-a_{2}^{+}a_{2}  \label{3}
\end{equation}%
which commutes with the $su(1,1)$ generators. The superalgebra $osp(2,2)$
might be constructed by extending $su(1,1)$ algebra with the fermionic
generators. These are given by 
\begin{equation}
V_{+}=f^{+}a_{2}^{+},\quad V_{-}=f^{+}a_{1},\quad W_{+}=fa_{1}^{+},\quad
W_{-}=fa_{2}  \label{4}
\end{equation}%
where $f^{+}$ and $f$ are fermions and they satisfy the anticommutation
relation 
\begin{equation}
\left\{ f,f^{+}\right\} =1\quad f=\sigma _{-},\quad f^{+}=\sigma _{+},\quad
ff^{+}-f^{+}f=\sigma _{0}.  \label{5}
\end{equation}%
The superalgebra $osp(2,2)$ can be constructed with the generators (\ref{1})
and (\ref{4}), as it is discussed in\cite{chen1}. The total number operator $%
J$ of the system and it is given by 
\begin{equation}
J=\frac{1}{2}N+\frac{1}{2}\left( f^{+}f-ff^{+}\right) .  \label{7}
\end{equation}%
The generators of the $osp(2,2)$ superalgebra satisfy the following
commutation and anticommutation relations: 
\begin{eqnarray}
\left[ J_{+},J_{-}\right] &=&-2J_{0},\quad \left[ J_{0},J_{\pm }\right] =\pm
J_{\pm },\quad \left[ J,J_{\pm }\right] =0,\quad \left[ J,J_{0}\right] =0 
\notag \\
\left[ J_{0},V_{\pm }\right] &=&\pm \frac{1}{2}V_{\pm },\quad \left[
J_{0},W_{\pm }\right] =\pm \frac{1}{2}W_{\pm },\quad \left[ J_{\pm },V_{\mp }%
\right] =V_{\pm },\quad \left[ J_{\pm },W_{\mp }\right] =W_{\pm },\quad \quad
\notag \\
\left[ J,W_{\pm }\right] &=&-\frac{1}{2}W_{\pm },\quad \left[ J,V_{\pm }%
\right] =\frac{1}{2}V_{\pm }\quad \left[ J_{\pm },V_{\pm }\right] =0,\quad %
\left[ J_{\pm },W_{\pm }\right] =0  \label{8} \\
\left\{ V_{\pm },W_{\pm }\right\} &=&J_{\pm },\quad \left\{ V_{\pm },W_{\mp
}\right\} =\pm J_{0}-J\quad \left\{ V_{\pm },V_{\pm }\right\} =\left\{
V_{\pm },V_{\mp }\right\} =0  \notag \\
\left\{ W_{\pm },W_{\pm }\right\} &=&\left\{ W_{\pm },W_{\mp }\right\} =0. 
\notag
\end{eqnarray}

The Hamiltonian of a physical system, with an underlying $osp(2,2)$
symmetry, has been expressed in terms of the operators of the corresponding
algebra.

\section{Transformation of the operators}

Transformation of the fermion boson representations of the $osp(2,2)$
algebra and its connection with the QES systems can be done by introducing
the following similarity transformation induced by the metrics%
\begin{equation}
S=(a_{2}^{+})^{-a_{1}^{+}a_{1}-\sigma _{+}\sigma _{-}}\quad \quad
T=(a_{2})^{a_{1}^{+}a_{1}+\sigma _{+}\sigma _{-}}.  \label{10}
\end{equation}%
These transformations lead to the single variable differential realizations
of the $osp(2,2)$ superalgebra. Using the operator $S$, the generators of $%
osp(2,2)$ takes the form:%
\begin{eqnarray}
J_{+}^{\prime } &=&SJ_{+}S^{-1}=a_{1}^{+}\quad J_{-}^{\prime
}=SJ_{-}S^{-1}=a_{1}(a_{2}^{+}a_{2}+a_{1}^{+}a_{1}+\sigma _{+}\sigma _{-}) 
\notag \\
J_{0}^{\prime } &=&SJ_{0}S^{-1}=\frac{1}{2}\left(
2a_{1}^{+}a_{1}+a_{2}^{+}a_{2}+1+\sigma _{+}\sigma _{-}\right) \quad
J^{\prime }=SJS^{-1}=\frac{1}{2}(-a_{2}^{+}a_{2}-\sigma _{-}\sigma _{+}) 
\notag \\
V_{+}^{\prime } &=&SV_{+}S^{-1}=\sigma _{+},\quad V_{-}^{\prime
}=SV_{-}S^{-1}=\sigma _{+}a_{1}  \label{17} \\
W_{+}^{\prime } &=&SW_{+}S^{-1}=\sigma _{-}a_{1}^{+},\quad W_{-}^{\prime
}=SW_{-}S^{-1}=\sigma _{-}(a_{2}^{+}a_{2}+a_{1}^{+}a_{1}+\sigma _{+}\sigma
_{-})  \notag
\end{eqnarray}%
The representations (\ref{17}) of $osp(2,2)$ can be characterized by a fixed
number $a_{2}^{+}a_{2}=-j-1.$Here $j$ takes integer or half-integer values.
Therefore the generators of the $osp(2,2)$ algebra can be expressed as
single variable differential equation in the Bargmann-Fock space and two
component polynomials of degree $j$ and $j+1$ form a basis function for the
generators of the $osp(2,2)$ algebra,$\qquad $%
\begin{equation}
P_{n+1,n}(x)=\left( 
\begin{array}{c}
x^{0},x^{1},\cdots ,x^{n+1} \\ 
x^{0},x^{1},\cdots ,x^{n}%
\end{array}%
\right)  \label{22}
\end{equation}

The general QES operator can be obtained by linear and bilinear combinations
of the generators of the $osp(2,2)$ superalgebra. Action of the QES operator
on the basis function (\ref{22}) gives us a recurrence relation, therefore,
the wavefunction is itself the generating function of the energy
polynomials. Under the transformation $T$ the generators of the $osp(2,2)$
algebra take the form%
\begin{eqnarray}
J_{+}^{\prime } &=&TJ_{+}T^{-1}=a_{1}^{+}(a_{1}a_{1}+a_{2}^{+}a_{2}+1+\sigma
_{+}\sigma _{-})\quad J_{-}^{\prime }=TJ_{-}T^{-1}=a_{1}  \notag \\
J_{0}^{\prime } &=&TJ_{0}T^{-1}=\frac{1}{2}\left(
2a_{1}^{+}a_{1}+a_{2}^{+}a_{2}+1+\sigma _{+}\sigma _{-}\right) \quad
J^{\prime }=TJT^{-1}=\frac{1}{2}\left( -a_{2}^{+}a_{2}-\sigma _{-}\sigma
_{+}\right)  \notag \\
V_{+}^{\prime } &=&TV_{+}T^{-1}=\sigma
_{+}(a_{2}^{+}a_{2}+a_{1}^{+}a_{1}+1+\sigma _{+}\sigma _{-})\quad
W_{+}^{\prime }=TW_{+}T^{-1}=\sigma _{-}a_{1}^{+}  \label{23} \\
V_{-}^{\prime } &=&TV_{-}T^{-1}=\sigma _{+}a_{1}\quad W_{-}^{\prime
}=TW_{-}T^{-1}=\sigma _{-}  \notag
\end{eqnarray}

This realization can also be characterized by $a_{2}^{+}a_{2}=-j-1$ . The
basis function of the realization is given by (\ref{22}).

\section{Solvability of the $E\otimes \protect\varepsilon $ Jahn-Teller
Hamiltonian}

It will be shown that our approach relatively very simple when compared
previous approaches. The Hamiltonian (\ref{eq:1}) can be expressed in terms
of the generators of the $osp(2,2)$:

\begin{equation}
H=2J_{0}+(\frac{1}{2}+2\mu )(2J-N)+2\kappa \left[ V_{+}+V_{-}+W_{+}+W_{-}%
\right] .  \label{29}
\end{equation}

The general trend to solve a differential equation quasi exactly is to
express the differential equation in terms of the generators of a given Lie
algebra having a finite dimensional invariant subspace and use the algebraic
operations. In the Bargmann-Fock space the Hamiltonian has two different
realization, under the transformation $S$ and $T$. The first transformation
by $S$ leads to the following one variable differential realization:%
\begin{equation}
H_{1}=\left( 2x\frac{d}{dx}-j+\sigma _{+}\sigma _{-}\right) -(\frac{1}{2}%
+2\mu )\sigma _{-}\sigma _{+}+2\kappa \left[ \sigma _{+}(1+\frac{d}{dx}%
)+\sigma _{-}(x+x\frac{d}{dx}-j-1+\sigma _{+}\sigma _{-})\right]  \label{30}
\end{equation}%
and the second realization can be obtained by transforming the Hamiltonian
by $T$ :%
\begin{equation}
H_{2}=\left( 2x\frac{d}{dx}-j+\sigma _{+}\sigma _{-}\right) -(\frac{1}{2}%
+2\mu )\sigma _{-}\sigma _{+}+2\kappa \left[ \sigma _{+}(\frac{d}{dx}+x\frac{%
d}{dx}-j+\sigma _{+}\sigma _{-})+\sigma _{-}(1+x)\right] .  \label{31}
\end{equation}%
The eigenvalue problem can be expressed as%
\begin{equation}
H\varphi (x)=E\varphi (x);\quad \varphi (x)=\left[ 
\begin{array}{l}
v_{n}(x) \\ 
\omega _{m}(x)%
\end{array}%
\right]  \label{32}
\end{equation}%
where $v_{n}(x)$ and $\omega _{m}(x)$ are polynomials of degree $n$ and $m$
respectively. The action of the $H_{1}$ on the basis function $\varphi (x)$
gives the following recurrence relation: 
\begin{eqnarray}
(2n-j+1-E)v_{n}+2\kappa (\omega _{m}+m\omega _{m-1}) &=&0  \notag \\
(2m-j-\frac{1}{2}-2\mu -E)\omega _{m}+2\kappa (v_{n+1}+(n-j)v_{n}) &=&0
\label{34}
\end{eqnarray}%
Similarly when the Hamiltonian $H_{2}$ act on the basis function we obtain
the recurrence relation: 
\begin{eqnarray}
(2n-j+1-E)v_{n}+2\kappa (m\omega _{m-1}+(m-j)\omega _{m}) &=&0  \notag \\
(2m-j-\frac{1}{2}-2\mu -E)\omega _{m}+2\kappa (v_{n}+v_{n+1}) &=&0
\label{35}
\end{eqnarray}

It is requiring that the determinant of these sets must be equal to zero
giving the compatibility conditions which establish the QES system.
According to the (\ref{22}) one can construct QES system if $n=m+1$. Here $m$
takes the values $m=0,\frac{1}{2},1,\cdots ,j.$ If $E_{j}$ is a root of the
recurrence relations (\ref{34}) or (\ref{35}) then the eigenfunction
truncates for a certain values of $j$ and $E_{j}$ belong to the spectrum of
the Hamiltonian. The initial conditions of the recurrence relation is given
by%
\begin{equation}
v_{m}=0;\text{for }j<m<1\text{ and }\omega _{m}=0\text{ for }j<m<0.
\label{37}
\end{equation}%
with this initial conditions solution of (\ref{34}) the gives us the
following relation for the energy when $j=0:$%
\begin{equation}
E=\frac{1}{4}\left( 5-4\mu \pm \sqrt{64\kappa ^{2}+(7+4\mu )^{2}}\right)
\label{38}
\end{equation}%
and for $j=1/2:$%
\begin{equation}
E=\frac{1}{4}\left( 3-4\mu \pm \sqrt{32\kappa ^{2}+(7+4\mu )^{2}}\right)
\quad E=\frac{1}{4}\left( 7-4\mu \pm \sqrt{64\kappa ^{2}+(7+4\mu )^{2}}%
\right)  \label{39}
\end{equation}%
The same energy eigenvalues can be obtained by using the recurrence relation
(\ref{35}). In this case eigenvalues shifted $E\rightarrow E-1$ and $j$
takes negative integer and half integer values.

\section{Conclusion}

It is well known that the exact solutions have a direct practical
importance. We have presented the quasi-exact solution of the generalized $%
E\otimes \varepsilon $ JT system. Our paper gives a unified treatment of
some earlier works. The method given here can be extended to other JT or
multi dimensional atomic systems. The basic features of our approach is to
construct $osp(2,2)$ invariant subspaces. Furthermore, we have presented two
different boson-fermion representations and two classes of one variable
differential realizations of $osp(2,2)$ algebra . In particular the solution
of $E\otimes \varepsilon $ JT system has been constructed.

The suggested approach can be generalized in various directions. Invariant
subspaces of the multi-boson and multi-fermion systems can be obtained by
extending the method given in this paper. The method given here can be
extended to other JT or multi dimensional atomic systems.

\end{document}